\documentclass[journal]{IEEEtran} 
\usepackage{cite}
\usepackage{graphicx}
\usepackage{amsmath}
\usepackage{grffile}
\usepackage{tabularx}

\graphicspath{{eps/}}

\ifCLASSOPTIONcompsoc
  \usepackage[caption=false,font=normalsize,labelfont=sf,textfont=sf]{subfig}
\else
  \usepackage[caption=false,font=footnotesize]{subfig}
\fi

\begin{document}
\title{Development of next-generation timing system for the Japan Proton Accelerator Research Complex}
\author{Fumihiko~Tamura, Hiroki~Takahashi, Norihiko~Kamikubota, Yuichi~Ito, Naoki~Hayashi 
\thanks{Manuscript received October 23, 2020}
\thanks{Fumihiko~Tamura, Hiroki~Takahashi, Norihiko~Kamikubota, Yuichi~Ito, and Naoki~Hayashi are with the J-PARC Center, 
JAEA \& KEK, 2-4 Shirakata, Tokai, Ibaraki 319-1195, Japan.
(e-mail: fumihiko.tamura@j-parc.jp)
}%
}

\maketitle
\thispagestyle{empty}
\begin{abstract}
A precise and stable timing system is necessary for high
intensity proton accelerators such as the J-PARC. The original timing
system, which was developed during the construction period of the
J-PARC, has been working without major issues since 2006.  After a
decade of operation, the optical modules, which are key components for
signal transfer, were discontinued already. Thus, the next-generation
timing system for the J-PARC has been developed and built. The new system is
designed to be compatible with the original system in terms of the
operating principle.  The new system utilizes modern high speed serial
communication for the transfer of the clock, reference trigger, and type code.
We describe the system configuration of the next-generation timing
system and current status.
\end{abstract}


\section{Introduction}
\IEEEPARstart{T}{he} Japan
Proton Accelerator Research Complex (J-PARC) \cite{TDR} is 
a high intensity proton accelerator facility, which includes
three accelerators, the 400~MeV linac, the 3~GeV rapid cycling
synchrotron (RCS), and the 30~GeV main ring synchrotron (MR).
The design output beam power of the RCS and MR is 1~MW
and 750~kW, respectively. The RCS delivers high intensity proton beams 
to the Material and Life Science Experimental Facility (MLF)
for generation of neutrons and muons and to the MR.
The MR beams are delivered to the neutrino experiment and the hadron hall 
by fast extraction (FX) and slow extraction (SX) schemes, respectively.
The beam commissioning of the J-PARC accelerators started in 2006. 
The beam intensity has been increased steadily with progress of 
the beam tuning and hardware upgrades. As of June 2020,
the RCS output beam power for the MLF user program is 600~kW,
and the MR beam power reached 510~kW and 51~kW for 
the neutrino experiment and the hadron experiments, respectively.

\begin{figure}[t]
 \centering
 \includegraphics[width=\linewidth]{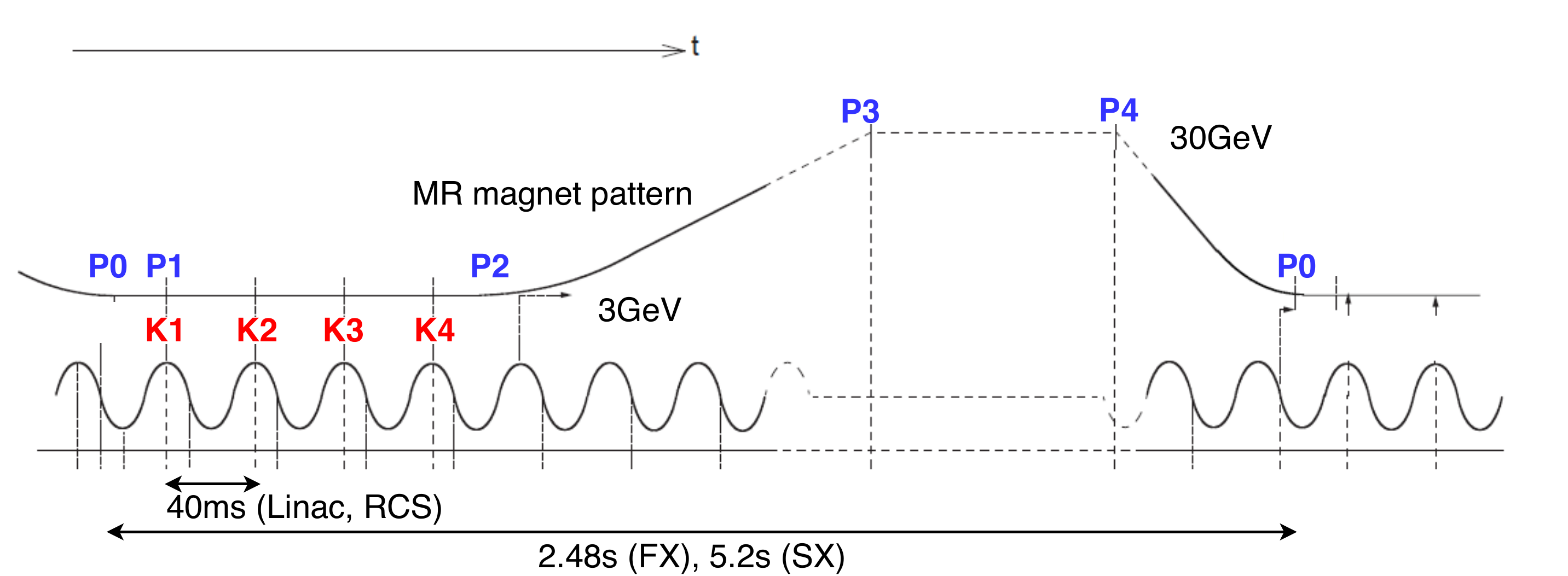}
 \caption{Operating cycle of the J-PARC accelerators.}
 \label{fig:MR_cycle}
\end{figure}

The operating cycle of the J-PARC accelerators is illustrated in Fig.~\ref{fig:MR_cycle}.
The linac and RCS are operated at the repetition rate of 25~Hz.
The MR cycle, i.e., the whole operating cycle of the J-PARC, is 2.48~s and 5.2~s for
the FX and SX modes, respectively. 
The MR cycle must be an integer multiple of the RCS period, 40~ms.
Four RCS beam is injected to the MR
at the K1 to K4 timing and the other beam is delivered to the MLF.
The beam parameters of the linac and RCS, such as the macro pulse width, the chopping
width, and the exciting pattern of the injection bump magnets,
 are different for the beam to the MLF and MR.
Therefore, different timing signals for the devices of the linac and RCS
are required during the MR cycle.

For high intensity proton accelerators such as the J-PARC, the beam loss
reduction is important to avoid the residual activation of the
components.  A precise and stable timing system, which generates
the triggers and gates for the accelerator devices, is necessary to realize
the precise beam control for minimizing the beam losses.
Typically, the jitter of the triggers must be less than 1~ns in the J-PARC.

At the J-PARC, two kinds of timing are defined; the scheduled timing 
and the synchronization timing. The scheduled timing is defined by
the programmed delay from the 25~Hz reference trigger sent from the central
control building (CCB). The synchronization timing is the timing signals
generated by the accelerator devices to synchronize the beams.
The linac chopper gate pulses are generated by the low level rf (LLRF)
control system of the RCS so that the injected intermediate beam
pulses into the RCS are centered to the rf buckets of the RCS.
The trigger for the extraction kicker of the RCS and the injection
kicker of the MR is also generated by the RCS LLRF control system.
Most of the accelerator devices in the J-PARC are operated
by the scheduled timing. We focus on the scheduled timing in this article.

The original timing system \cite{tamura:icalepcs03,tamura:pac05-timing}  
was developed during the construction period
of the J-PARC. The system has been working without major problems since
the start of the beam operation in 2006. However, the optical modules,
which are key components in the timing system for signal transfer,
have been discontinued already. We decided to develop the next-generation
timing system in 2016 and the new modules were successfully deployed
in 2020. 

In this article, we describe the operating principle of the schedule timing
and the configuration of the original timing system. The configuration of the next-generation
timing system is presented.  
Also, the test results and the current status
are described.

\section{Original timing system}
\subsection{Operating principle of the scheduled timing}

\begin{figure}[t]
 \centering
 \includegraphics[width=\linewidth]{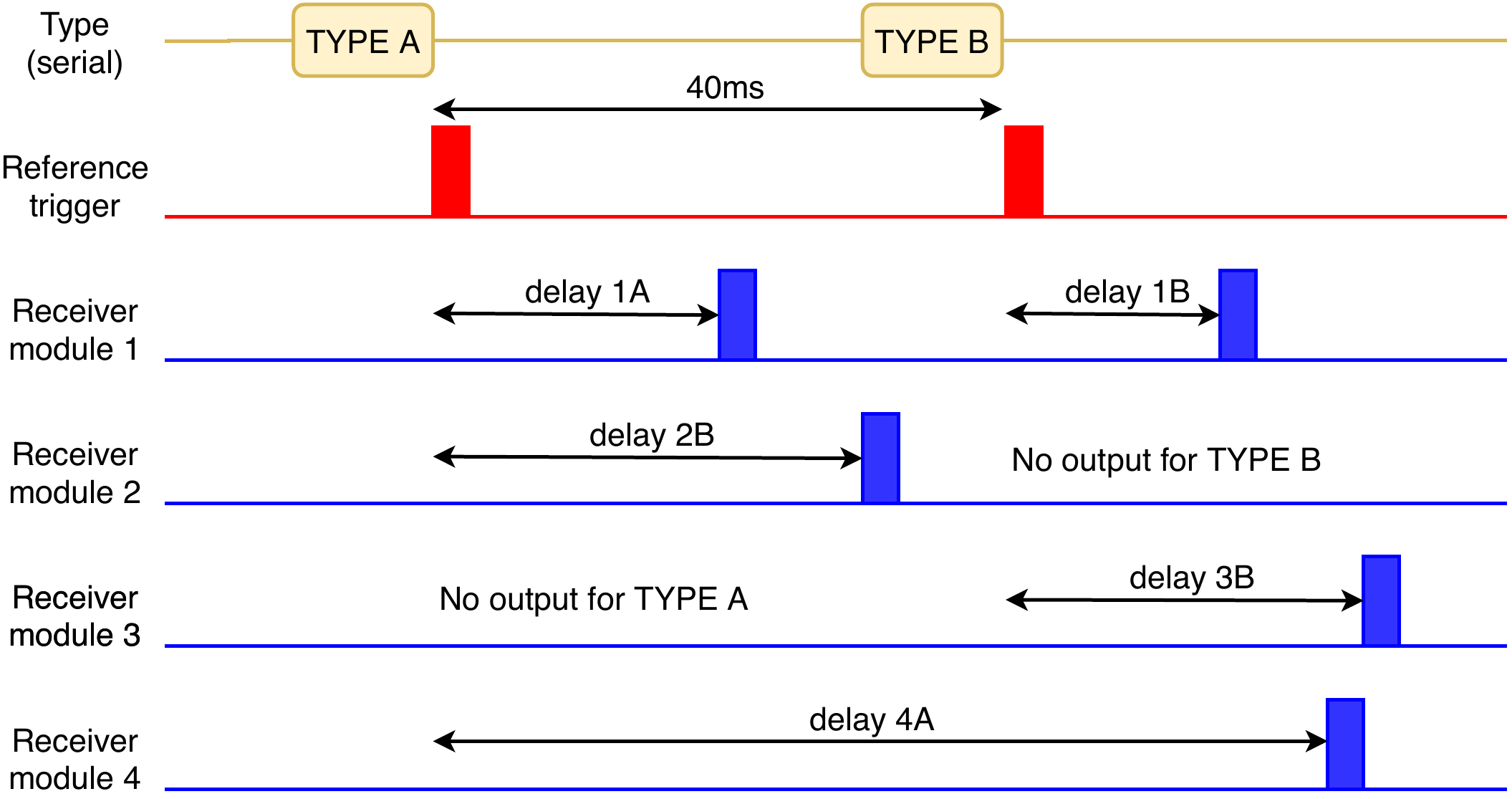}
 \caption{Operating principle of the scheduled timing.}
 \label{fig:principle_scheduled_timing}
\end{figure}

Operating principle of the scheduled timing is shown in
Fig.~\ref{fig:principle_scheduled_timing}. 
The 25~Hz reference trigger and the type code, which corresponds to the operation type
of the accelerators for the next linac and RCS cycle,
are sent from the CCB to the receiver modules in three accelerators
and three experimental buildings. The type code is sent prior to the 25~Hz
reference trigger.

The timing receiver module has a look-up table (LUT) with the type code
as address. The behavior of the receiver after the next reference trigger is 
defined by the control word from the LUT. Three behaviors are defined
as follows.
First, the receiver outputs a trigger with the programmed delay.
In the figure, the receiver module 1 and 2 output triggers
at the delay 1A and 2A for TYPE A, respectively. For TYPE B,
the module 1 and 3 generate the triggers the similar manner.
Second, the output can be suppressed for the next 40~ms.
The outputs of the receiver modules 2 and 3 are suppressed
for TYPE B and A, respectively.

The delay value can be set longer than the 40~ms, as the delay 4A of
the receiver module 4
for TYPE A in the figure, while normally the delay counter is reset
with the next recerence trigger. With the special control word for TYPE B, 
the delay counter continues beyond the next reference trigger.

The type sequence, of which length is same as the MR cycle, represents
the operation of the accelerators every 40~ms. If the end code is picked up from
the type memory, the sequence start from the beginning of the type memory.
In the case of the jump code, the next sequence starts with different 
type memory indicated by the jump code. When the type memory
is switched manually by a software, the switching happens at the end
or jump code. This guarantees the MR cycle is fixed during a run.
By switching of the type
sequence as well as the modification of the LUTs, various operation
of the J-PARC accelerator is realized.

The operating principle is similar to the modern event timing
system such as the MRF (Micro Research Finland) 
timing system \cite{MRF:timing}, while the event, i.e., the type code,
is sent in a fixed repetition of 25~Hz in the J-PARC timing system.

\begin{figure}[tb]
\centering
\includegraphics*[width=0.7\linewidth]{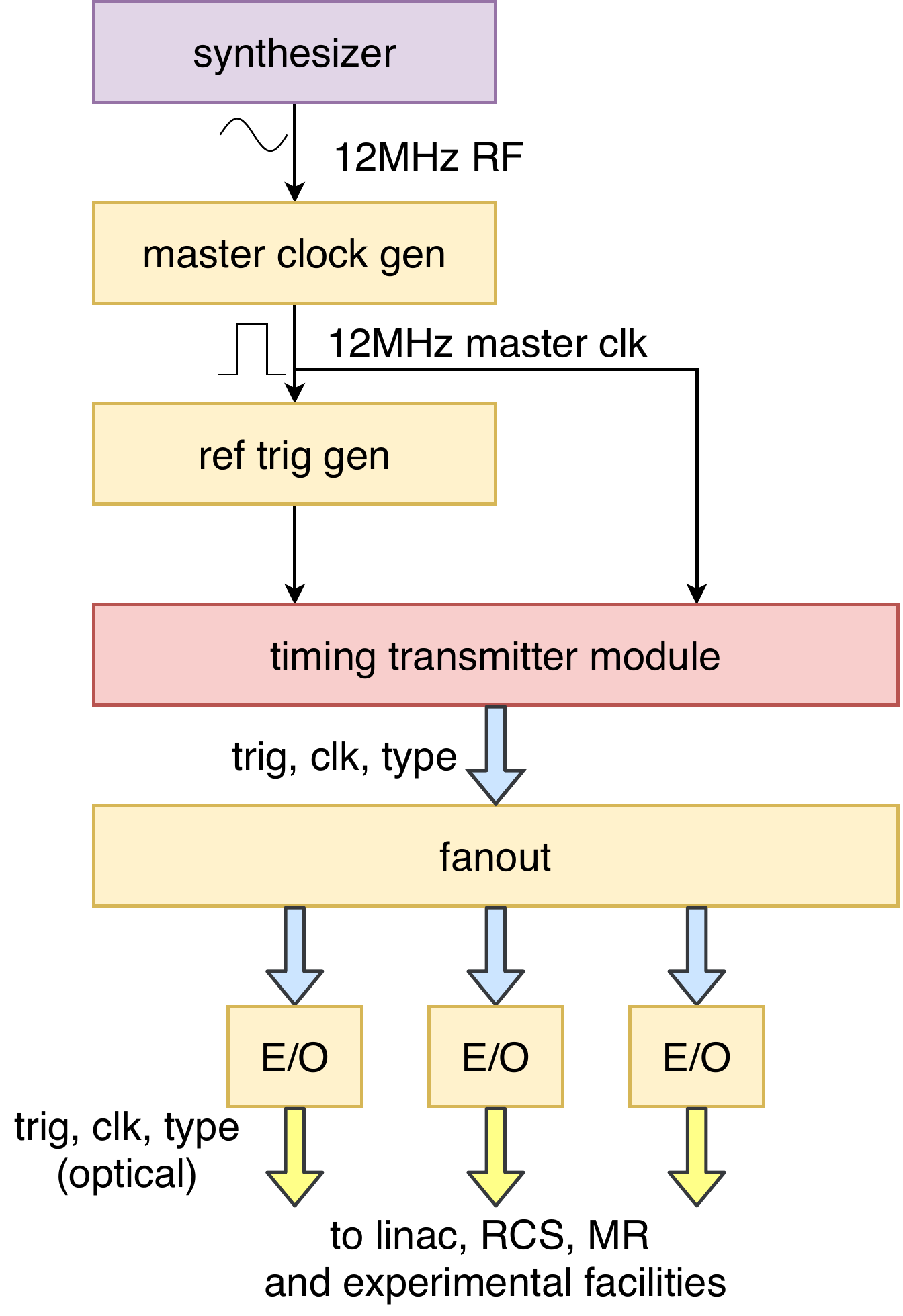}
\caption{Signal flow of the timing transmitter station in the CCB.}
\label{fig:transmitter_block}
\end{figure}

\begin{figure}[tb]
\centering
\includegraphics*[width=0.8\linewidth]{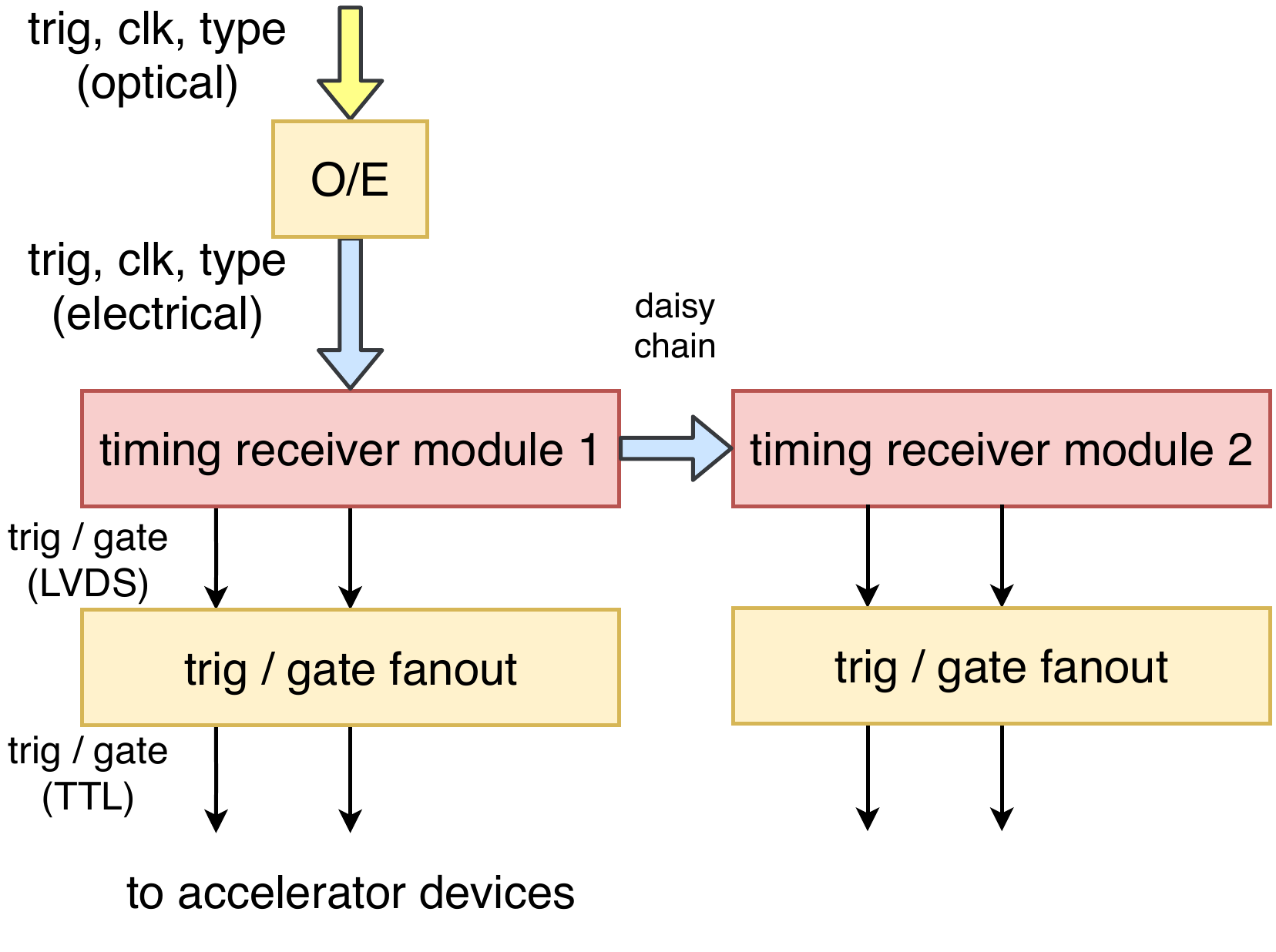}
\caption{Signal flow of the timing receiver station in the facilities.}
\label{fig:receiver_block}
\end{figure}

\subsection{Hardware configuration}

Figure~\ref{fig:transmitter_block} shows the signal flow of 
the timing transmitter station in the CCB. The 12~MHz master clock
is generated by the master clock generator referring the rf signal 
from a high precision synthesizer. By counting the master clock,
the 25~Hz reference trigger is generated. The timing transmitter module,
which is the core of the J-PARC scheduled timing, outputs
the serialized type code according to the programmed type
sequence. Through the fanout module and the E/O (electrical/optical) converter,
the three optical signals, the 25~Hz reference trigger, the master clock, and the type code,
are sent to the facilities of the J-PARC.

The signal flow is a star configuration. All receiver modules in the facilities
receive the same three signals. If necessary, a combination of the O/E, fanout, and 
E/O can be used to redistribute the signals at each facility.

The type code is a 32 bit word, with the first
bit representing special information. Eight bits each are allocated to the linac,
RCS, and MR. The remaining seven bits are reserved.

The signal flow of the timing receiver station is illustrated in Fig.~\ref{fig:receiver_block}.
The optical signals of the 25~Hz reference trigger, the master clock, and the type code
are converted by the O/E module to the electric signals 
and they are lead to the timing receiver module.
The signals can be forwarded to another receiver modules with a daisy chain.

A 96~MHz clock is generated by a phase
lock loop (PLL) in the receiver module referring the 12~MHz master clock.
The delay counter in the module uses the 96~MHz clock.
The receiver module retrieves the control word from the LUT by taking the
specified 8 bits from the 32 bit type code. The behavior of the receiver
module for the next 25~Hz reference trigger is defined by the control word, as described
in the previous subsection. The delay value is 24 bit, therefore the delay
counter can count up to 174~ms. The signal level of the receiver module
is LVDS, and the following trigger and gate fanout module converts
to the signal level for the accelerator devices.

The timing transmitter and receiver modules are VME modules,
and the other modules are NIM module.
In the linac and RCS, the reflective memory (RM) networks are configured
together with the timing system \cite{takahashi:epac08}.
The RM network enables fast rewriting the LUTs and sharing the beam tag
information. A synchronized data acquisition at 25~Hz is realized with the RM
network for the RCS and the fast data acquisition is indispensable
for the commissioning and operation of the RCS \cite{takahashi2015application}.

We note here that the timing of the J-PARC is based on the 12~MHz
master clock generated referring the high precision synthesizer,
i.e., the accelerator cycle is not synchronized to the AC power line,
which has a frequency variation in the order of 0.1\%.
The jitter of the generated triggers is very low, less than 1~ns.
By the accurate timing system and the digital LLRF control system
of the RCS, the very low beam timing jitter of 1.7~ns,
which satisfies the requirement of the fermi chopper spectrometer \cite{itoh:nim12-fermichopper}
in the MLF,
is achieved \cite{tamura:icaplepcs09,tamura:nima11}.

\subsection{Demand of the next-generation timing system}

The original timing system started its operation in 2006. Since then, the system
 has been working well without major problems for more than ten years.
However, the optical device (Finisar v23826) in the E/O and O/E modules,
which are the key modules for the signal transfer, 
has already been discontinued, and there are no successor devices.
While we have an amount of spare modules, it will be soon difficult
to maintain the original timing system. Therefore, we started to
consider a next-generation timing system in 2016.

The number of the timing receiver modules is in the order of 200 \cite{kamikubota:icalepcs15}.
It is not realistic to replace all of the original modules with the next-generation
system at the same time. Therefore, the new system is designed to be compatible with
the original system and the operating principle is employed as is.
From this viewpoint, the popular timing solutions such as the MRF timing system \cite{MRF:timing}
or the White Rabbit \cite{Liminski:WR} are not considered to be employed.

\begin{figure}[tb]
\centering
\subfloat[Original system.\label{fig:timing_before}]{
 \includegraphics[width=\linewidth]{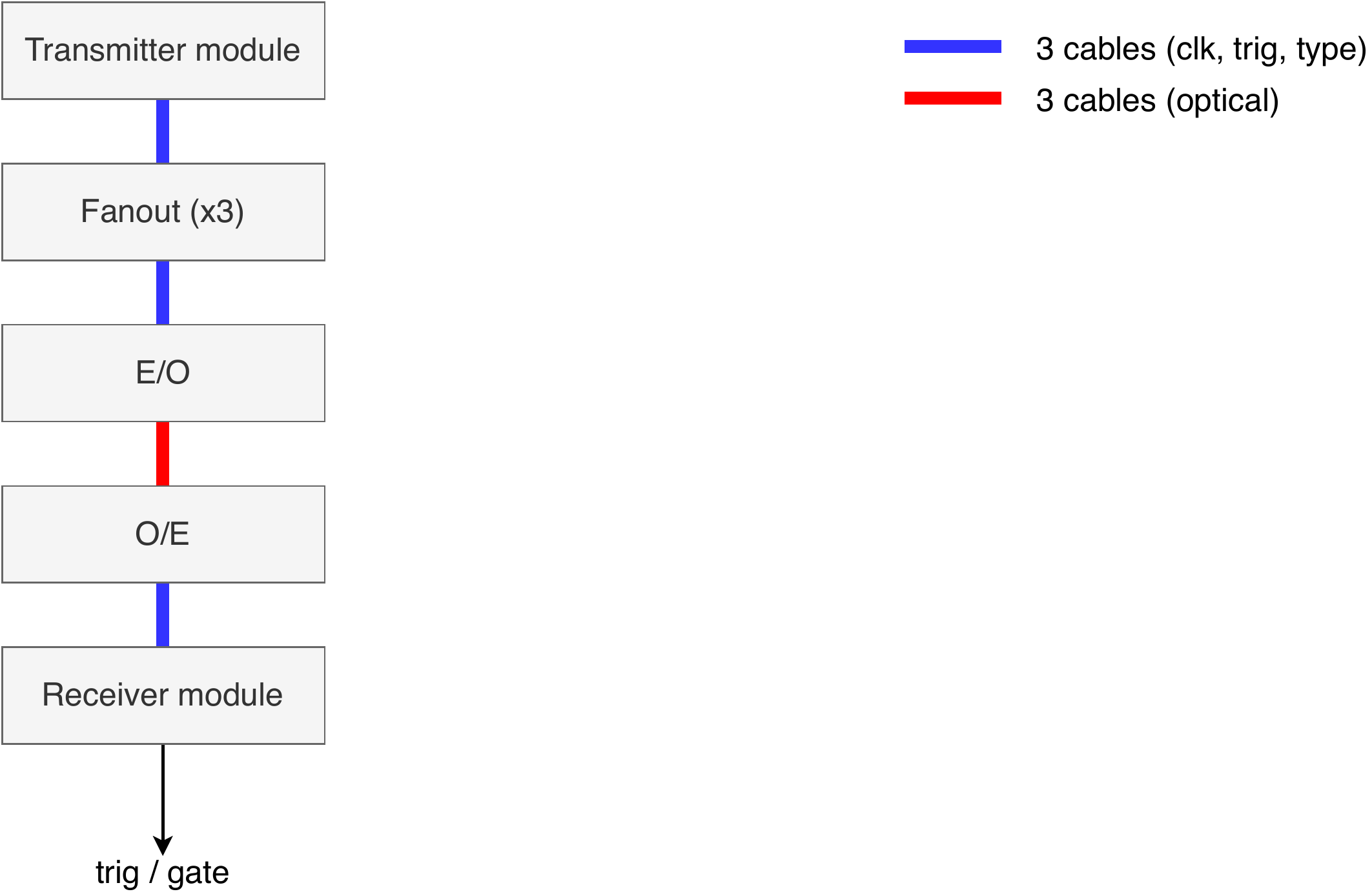}
 }
\vspace*{2ex}
\subfloat[Next-generation system.\label{fig:timing_after}]{
 \includegraphics[width=\linewidth]{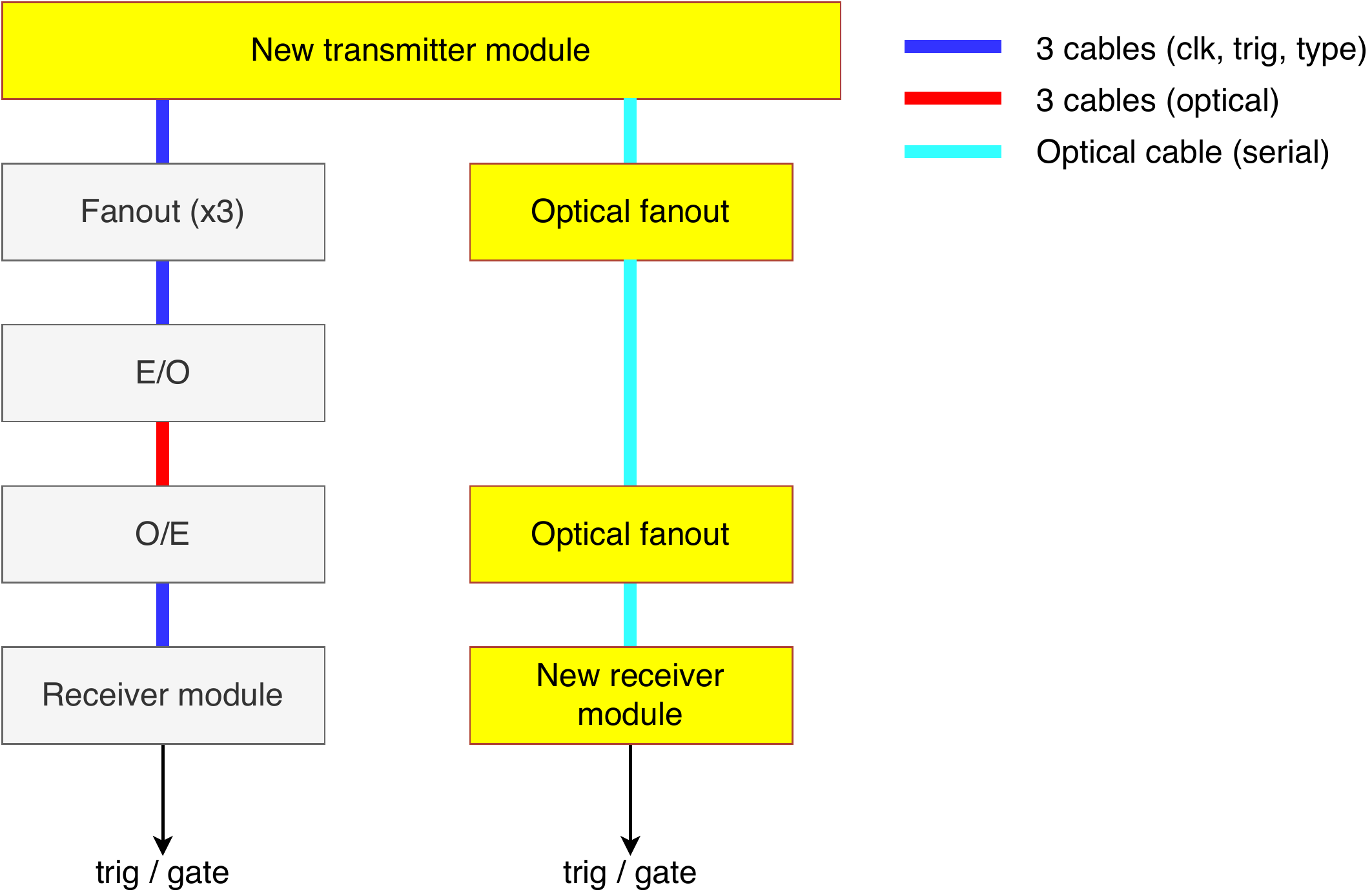}
 }
\caption{Comparison of the system configurations. The blue and red line
represent the set of three electrical and optical cables, respectively.
The light blue line represents a single optical cable.}
\label{fig:timing_before_after}
\end{figure}

After a decade of the operation, we found the following issues.
It has been an infrastructure burden that
the original system requires three optical or metal 
cables for transfer of three signals, the 25~Hz reference trigger, the master clock,
and the type code, as shown in Fig.~\ref{fig:timing_before_after}\subref{fig:timing_before}.
Also, the metal cable between the O/E module and the receiver module
may pick up noises from the pulsed power supplies or the high intensity beams,
even with short metal cables.
The receiver module may suffer from the noises; 
We have experienced several hang-ups of the receiver module due to the noises.

A set of VME and NIM crates are necessary, which require an amount of space and cost,
 for the timing receiver station of the original system. This is an important issue
especially for the MR, where the hardware extension is still ongoing and
the new deployment of the receiver station is often required.

The next-generation system has been designed considering these issues.

\section{Next-generation timing system}

\begin{table*}[tb]
 \caption{Protocol of the high speed serial communication}\label{tab:serial_protocol}
\subfloat[Format\label{tab:serial_format}]{
{\scriptsize
 \begin{tabularx}{\textwidth}{|p{2.0em}||X|X|X|X|X|X|X|X|X|X|X|X|}
 \hline
 & 1& 2& 3& 4& 5& 6&7& 8&9&10&11&12\\
 \hline
 seq & K28.5& E & D(p1)& D(p2)& D(p3)& D(p4)&  D(p5)& D(p6) &D(p7)&D(p8)&D(p9)&D(p10)\\
 \hline
 \end{tabularx}}}

\subfloat[Null event. E(null)=D0.0 (0x00)]{
{\scriptsize
 \begin{tabularx}{\textwidth}{|p{2.0em}||X|X|X|X|X|X|X|X|X|X|X|X|}
  \hline
 & 1& 2& 3& 4& 5& 6&7& 8&9&10&11&12\\
 \hline
 Null & K28.5& E(null) & D0.0 & D0.0 & D0.0 & D0.0 &  D0.0 & D0.0& D0.0 & D0.0 &  D0.0 & D0.0\\
 \hline

 \end{tabularx}}}

\subfloat[Trigger event. E(trig)=D1.0 (0x01)]{
{\scriptsize
 \begin{tabularx}{\textwidth}{|p{2.0em}||X|X|X|X|X|X|X|X|X|X|X|X|}
 \hline
 & 1& 2& 3& 4& 5& 6&7& 8&9&10&11&12\\
 \hline
 Trig & K28.5& E(trig) & D0.0 & D0.0 & D0.0 & D0.0 &  D0.0 & D0.0& D0.0 & D0.0 &  D0.0 & D0.0\\
 \hline
 \end{tabularx}}}

\subfloat[Type event. E(type)=D2.0 (0x02)]{
{\scriptsize
 \begin{tabularx}{\textwidth}{|p{2.0em}||X|X|X|X|X|X|X|X|X|X|X|X|}
 \hline
 & 1& 2& 3& 4& 5& 6&7& 8&9&10&11&12\\
 \hline
 Type & K28.5& {\tiny E(type)} & type1 & type2 & type3 & type4 &  D0.0 & D0.0& D0.0 & D0.0 &  D0.0 & D0.0\\
 \hline
 \end{tabularx}}}

\subfloat[S event. E(S)=D3.0 (0x03)]{
{\scriptsize
 \begin{tabularx}{\textwidth}{|p{2.0em}||X|X|X|X|X|X|X|X|X|X|X|X|}
 \hline
 & 1& 2& 3& 4& 5& 6&7& 8&9&10&11&12\\
 \hline
 S & K28.5& E(S) & D0.0 & D0.0 & D0.0 & D0.0 &  D0.0 & D0.0& D0.0 & D0.0 &  D0.0 & D0.0\\
 \hline
 \end{tabularx}}}

\subfloat[S count event. E(Scnt)=D4.0 (0x04)]{
{\scriptsize
 \begin{tabularx}{\textwidth}{|p{2.0em}||X|X|X|X|X|X|X|X|X|X|X|X|}
 \hline
 & 1& 2& 3& 4& 5& 6&7& 8&9&10&11&12\\
 \hline
 Scnt & K28.5& {\tiny E(Scnt)} & Scnt1 & Scnt2 & Scnt3 & Scnt4 &  D0.0 & D0.0& D0.0 & D0.0 &  D0.0 & D0.0\\
 \hline
 \end{tabularx}}}

\subfloat[Trigger count event. E(Tcnt)=D5.0 (0x05)]{
{\scriptsize
 \begin{tabularx}{\textwidth}{|p{2.0em}||X|X|X|X|X|X|X|X|X|X|X|X|}
 \hline
 & 1& 2& 3& 4& 5& 6&7& 8&9&10&11&12\\
 \hline
 Tcnt & K28.5& {\tiny E(Tcnt)} & Tcnt1 & Tcnt2 & Tcnt3 & Tcnt4 &  D0.0 & D0.0& D0.0 & D0.0 &  D0.0 & D0.0\\
 \hline
 \end{tabularx}}}

\end{table*}

\subsection{System configuration}

The configurations of the next-generation timing system is shown
in Fig.~\ref{fig:timing_before_after}\subref{fig:timing_after}. In the figure, the blocks
with yellow color are the newly developed modules.

The new timing transmitter module output a serialized optical
signal, which contains the information of the 25~Hz reference trigger,
12~MHz clock, and the type code. The new transmitter also outputs 
the same three electrical signals as the original system,
so that the existing signal distribution networks and 
the timing receiver stations as they are. This allows for
a gradual transition to the next-generation system, 
while coexisting with the original timing modules.

A SFP (small form factor pluggable) optical transceiver,
which is widely used today, is employed as the optical component
for high speed serial communication.
The optical signal from the new transmitter module
is distributed to the new receiver modules in the facilities
via the newly developed optical fanouts.
Contrary to the original system, the optical signal
is directly input to the new receiver module.
It is expected that the malfunctions caused by the noises will
be mitigated. Similar to the original receiver module,
the optical signal can be forwarded to the another module forming a daisy
chain.

The new receiver module decodes and extracts the 12~MHz clock, 25~Hz
recerence trigger and type code from the received serial signal.  After decoding
the three signals, the operation is identical to that of the original
receiver module.

The new transmitter and receiver modules are implemented as VME
modules. The arrangement of the VME registers is the same as the
existing module as much as possible, so that the development of the
driver software is simple.  The PLC type receiver module is also
developed. The PLC receiver module is designed to output the triggers
and gates at the required voltage level form the accelerator devices
without the NIM modules. The PLC module is planned to be used extensively
in the MR, which has many restrictions on the arrangement of the control
devices.

\begin{figure}[tb]
\centering
\includegraphics[width=0.9\linewidth]{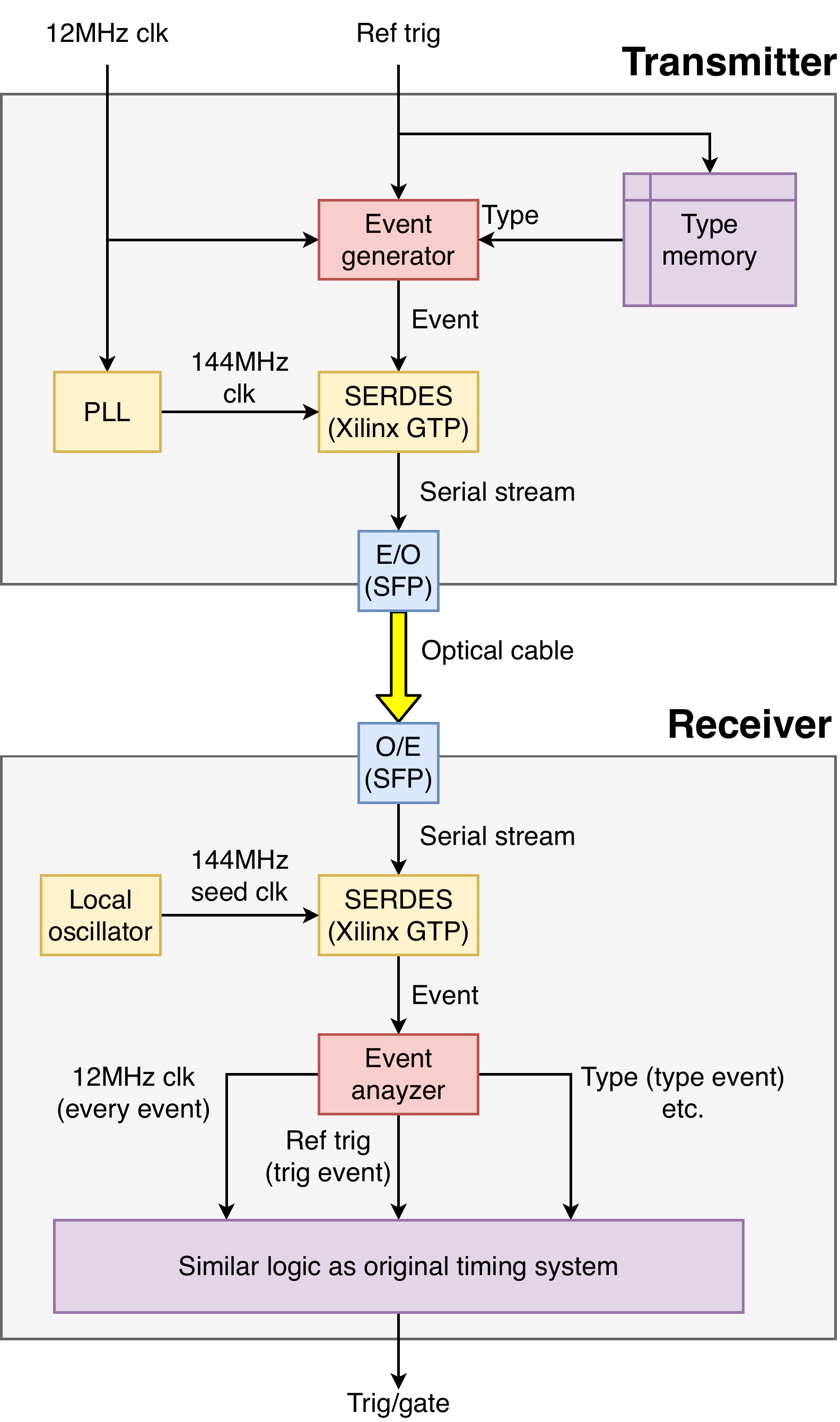}
\caption{Configuration of the high speed serial communication.}
\label{fig:new_timing_GTP}
\end{figure}

\subsection{High speed serial communication}

The configuration of the high speed serial communication in
the next-generation timing system is shown in Fig.~\ref{fig:new_timing_GTP}.

The new transmitter module generates an event, which is the payload of
the high speed serial communication, every 12~MHz clock.  Using the GTP \cite{xilinx:GTP}, a
high-speed communication transceiver in Xilinx FPGAs up to 3.2~Gbps, the generated
events are serialized and sent via SERDES.

The new system is a kind of the event timing systems.
The actual protocol of the serial communication to be sent is shown in
Table~\ref{tab:serial_protocol}.  In the 8B/10B conversion
\cite{8B10B_patent}, there is the D-character, which is the actual 8-bit
data, and the K-character with a special meaning.  K28.5 is used in
8B/10B as a comma character to indicate the delimitation of a data
sequence.  An event is sent with a comma followed by data, E, that
represents the meaning of the event, as shown in
Table~\ref{tab:serial_protocol}~\subref{tab:serial_format}.  The third
through twelfth word can contain a total of 80~bits of data, if
necessary.  An event is a sequence of 12 consecutive 8 bits, which is
converted to 8B/10B and transmitted at a transmission rate of 1.44~Gpbs.

When idle, the null event is constantly being sent. When the reference
trigger is input to the new transmitter module, the trigger event is
sent.  In the type event, the third through sixth in the sequence
contain type~1 through type~4 corresponding to mode~1 through mode~4,
where the mode~2, mode~3, and mode~4 are for the linac, RCS, and MR,
respectively. The type~1 for the mode~4 is reserved.
In addition, S and S count values and the trigger count values that are
transmitted by the original timing system are defined and transmitted as
events, where S indicates the beginning of the MR cycle and the trigger
count is the number of the reference 25~Hz triggers sent.
Since the events can be defined in 256 ways, additional event types can
be added and information can be sent as needed in the future.

\begin{figure}[tb]
\centering
\includegraphics[width=\linewidth]{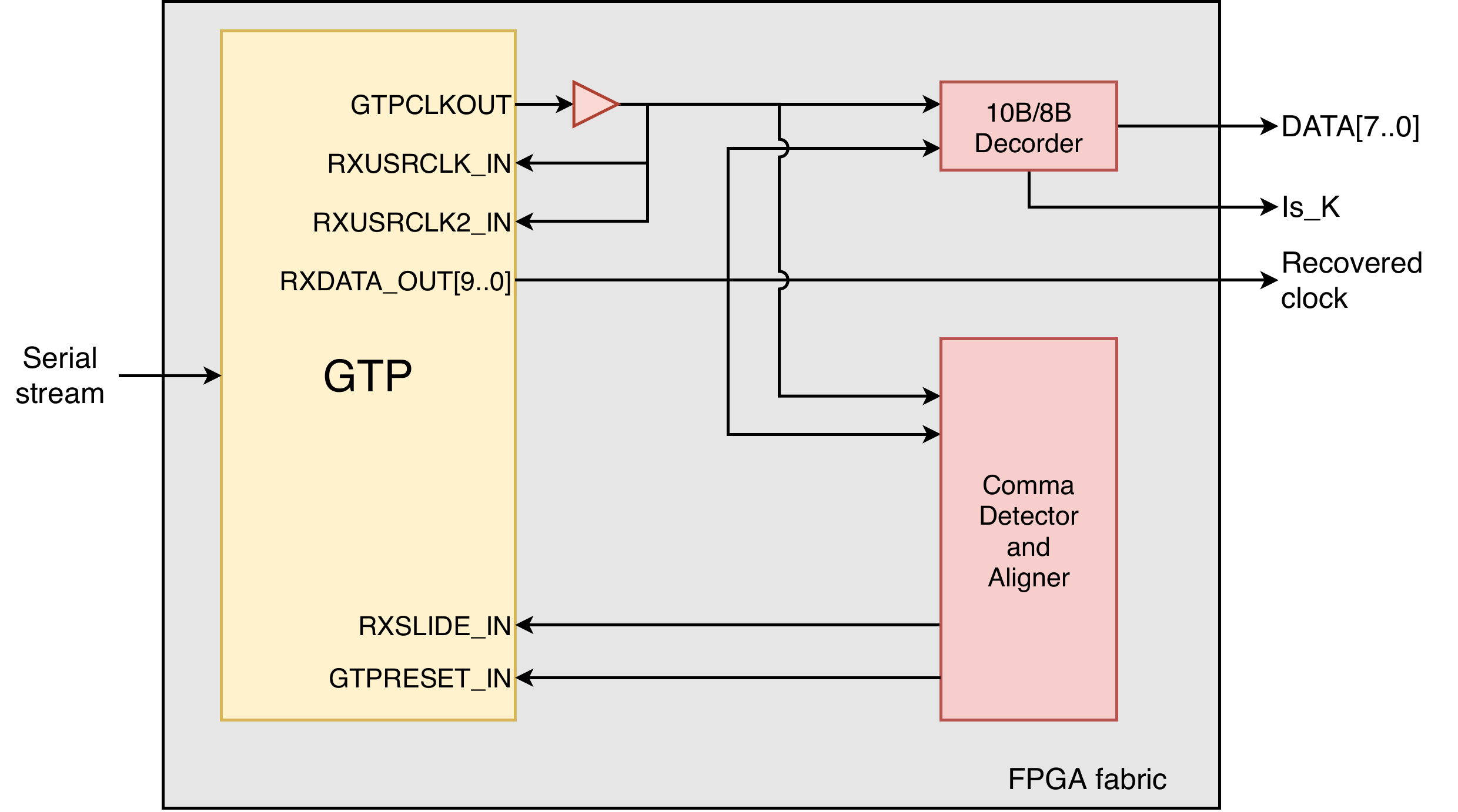}
\caption{Simplified block diagram of the fixed latency circuit.}
\label{fig:fixed_latency_block}
\end{figure}

\begin{figure}[tb]
\centering
\includegraphics[width=\linewidth]{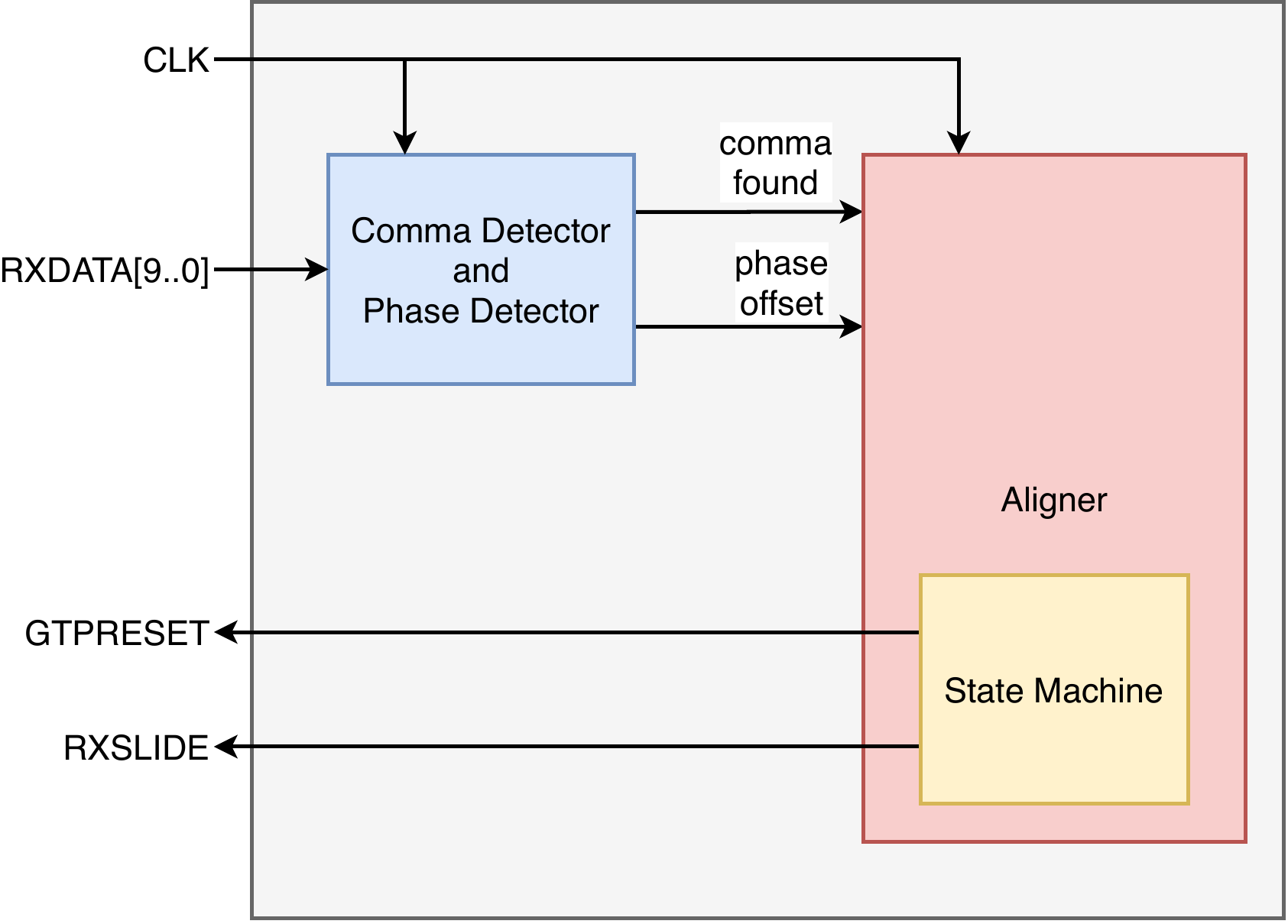}
\caption{Simplified block diagram of the comma detector and aligner.}
\label{fig:CDA_block}
\end{figure}

The serialized data is converted to the optical signal by the SFP
module, and delivered to the receiver module via the optical fanout
and the optical cables.

The 144~MHz clock is decoded and the data is extracted using clock data
recovery (CDR) by the GTP on the new receiver module.  The decoded
144~MHz clock is perfectly synchronized to the clock used in the
transmitter module. The received data is input to the event analyzer,
The event is sent at 12~MHz repetition and the event analyzer generate
12~MHz clock using the events.
 The reference trigger is decoded in the trigger event, and the
type code is decoded in the type event and supplied to the logic at the
latter stage. The other events are also decoded.  The logic of the
latter stage is basically the same as that of the original receiver
module and performs the same operation.

Thus, high speed serial communication makes it possible to transmit
three types of signals over a single optical cable.  It is important to
note here that the GTP transceiver has slightly different delays at each
power up.  The delay change is one period of the clock input to the
SERDES and is approximately 7~ns for the 144~MHz clock.  This is not an
acceptable change for the scheduled timing system. The recovered clock
is generated by dividing the high speed serial clock, therefore
several different phases are generated at each power up.
To realize a fixed latency, a circuit configuration similar to that in
Ref. \cite{giordano:tns11} is required.

According to the reference, the comma detector and aligner (CDA)
and the 10B/8B decoder for the receiver
are implemented in the FPGA fabric, as shown in Fig.~\ref{fig:fixed_latency_block}.
The most important part of the implementation is the CDA functionality.
The simplified block diagram of the CDA is shown in Fig.~\ref{fig:CDA_block}
The comma detector and phase detector block detects 
the phase misalignment by using the 10 bit RXDATA
and the recovered clock. When the K28.5 comma is found, the comma found
signal is asserted and the 4-bit phase offset information is sent
to the aligner block. In the aligner, a state machine is running
the following manner. If the phase offset in an even number of the unit intervals (UIs),
the RXSLIDE signal, which generates the 2~UI
phase shift of the recovered clock, is asserted for the necessary times.
If it is an odd number, it is impossible to adjust the phase by the RXSLIDE.
In this case the GTPRESET is asserted and the startup
process of the GTP runs again. The reset is repeated until the misalignment
becomes an even number.
With the implementation, the delay is fixed at all power ups.

\begin{figure}[tb]
\centering
\includegraphics[width=\linewidth]{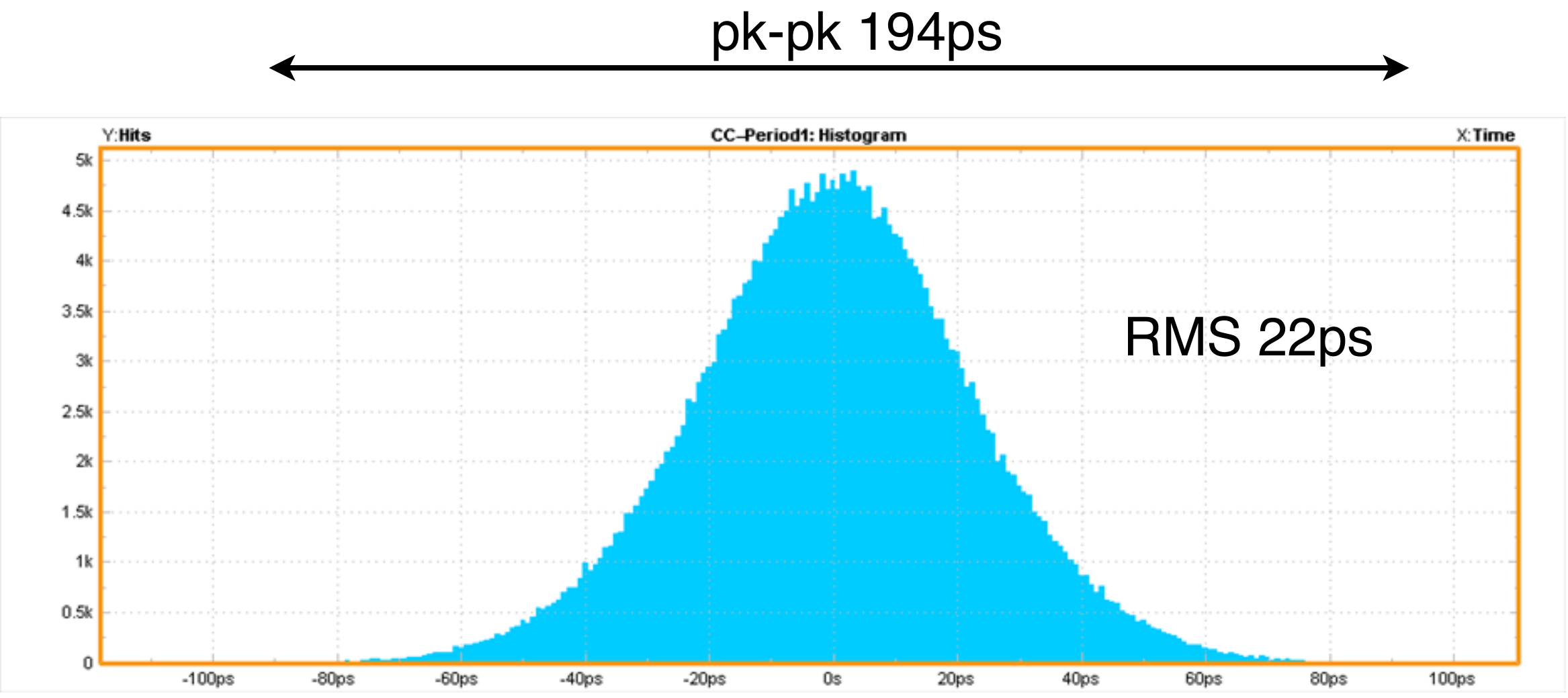}
\caption{Jitter measurement of the recovered 12~MHz clock by Tektronics DPOJET.}
\label{fig:measured_jitter_12MHz}
\end{figure}

The jitter of the 12~MHz clock decoded by the new receiver module is
important, because the 12~MHz clock is used not only as the reference
for the 96~MHz delay counter clock but also for generating the system clock
for the LLRF control system of the synchrotron.  It has been examined
the DPOJET of a Tektronics oscilloscope.  The measurement result is
shown in Fig.~\ref{fig:measured_jitter_12MHz}.  The jitter is 22~ps for
RMS and just under 200~ps for pk-pk, 
and sufficiently low to meet the requirements.
The jitter of the trigger output is similar.

Since the next-generation system utilizes the high speed serial communication,
the additional delay in the order of 250~ns compared to the original system
is introduced. The additional delay is fixed value, therefore it is not
problematic.

\begin{figure}[tb]
\centering
\includegraphics[width=\linewidth]{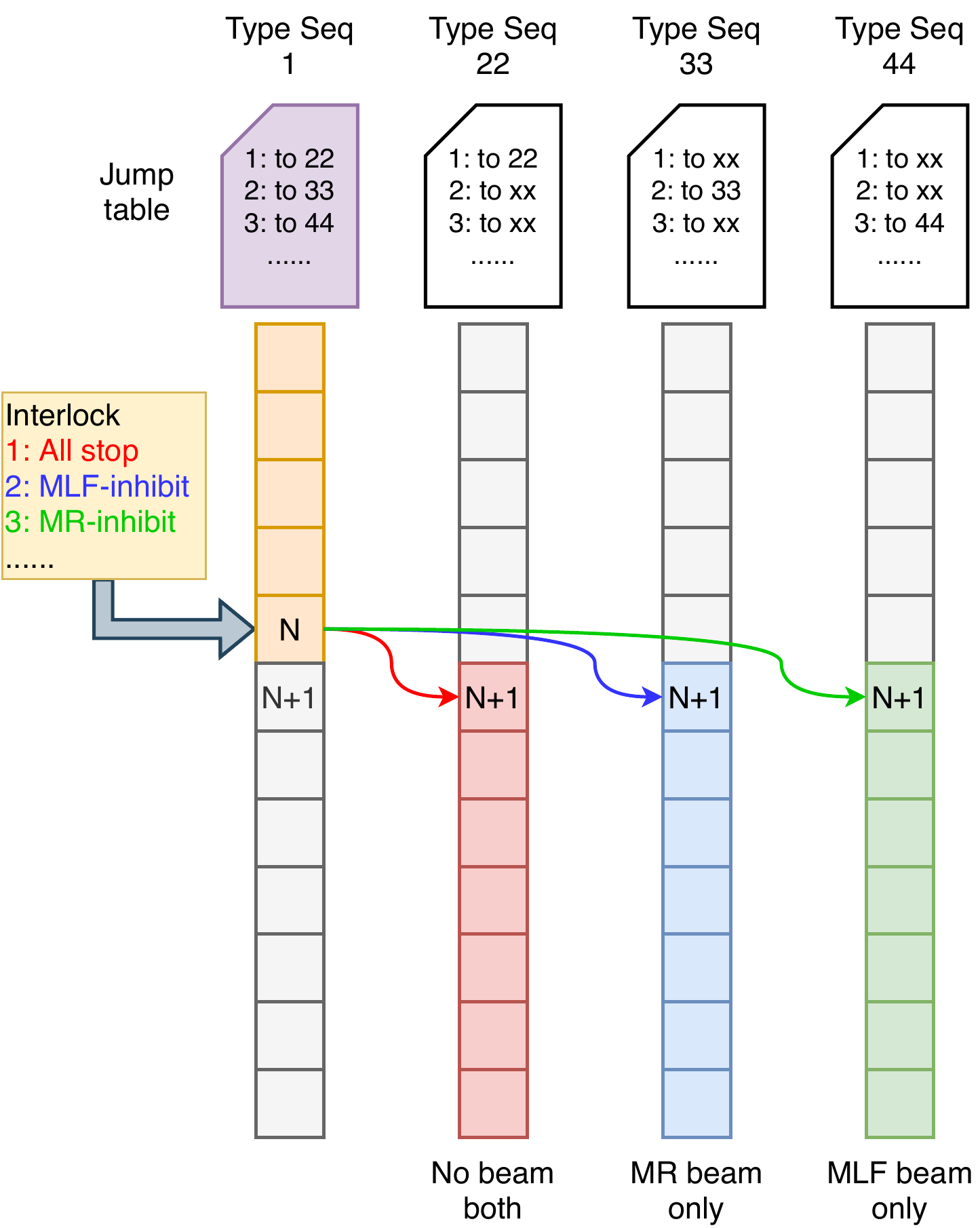}
\caption{Operating principle of the shift-jump function.}
\label{fig:shift_jump}
\end{figure}

\subsection{New features of the transmitter module}

As described in the previous subsection, the behavior of the new
receiver module is basically the same as the original receiver module,
while new features are implemented in the new transmitter module.

The length of the type memory in the original and new transmitter module 
is 512. the length corresponds to 20.48~s at the 25~Hz operation of the linac and
10.48~s at the 50~Hz operation, which is required for the future
extension with the transmutation experimental facility (TEF)\cite{sasa:2015design}.
In both cases, the lengths can cover the MR cycle, up to 5.2~s.
The number of the banks of the type memory is increased from 64
in the original module to 1024 in the new module. 
The linac will be operated at 50~Hz to inject the beams to the TEF
and RCS alternatively. The extension of the memory bank is necessary
to manage the increase of the combination of the operating patterns
with the TEF.

For high intensity accelerators such as the J-PARC, a machine protection system \cite{Schmidt:2206739}
(MPS), which quickly stops the beam at machine failures, is indispensable
to avoid damages of accelerator components due to the beam.
A fast MPS was developed for the J-PARC \cite{chiba:icalepcs03,sakaki:JAEA-REPORT2004}.
The J-PARC MPS can block the linac beam as quick as a few tens of $\mu$s,
by stopping the rf power to the RFQ (radio frequency quadrupole) at the beginning
of the linac. The interlock signals are sent to the MPS main unit in the CCB, and
the beam stop signal is generated and sent to the front-end of the linac.
The failures that trigger the MPS are classified into three
types, All-stop, MLF-inhibit \cite{sakai2014operation}, and MR-inhibit.
The All-stop interlock, which stops the beams for both of the MLF and MR, is asserted with the failures
at the linac and RCS, for example, the high level beam losses and the power supply failures.
If a failure happens in the MLF, the MLF-inhibit interlock is raised, where the linac beams
for the MLF are blocked and the beams for the MR are delivered to the RCS and MR.
The process is similar to the MR-inhibit interlock.

The beam triggers for the beam instrumentation devices such as the beam position monitors
and the current monitors should be stopped during the interlock, to avoid the loss of
the measured beam signals, which can be used for the postmortem analysis of the beam
or for the investigation of the machine failures. The beam triggers for the MR are
suppressed by switching of the type sequence by software
and that for the linac and RCS are stopped by
modifying the LUT of the receiver modules via the RM networks.
A new function, called the shift-jump function, is implemented in the new
transmitter module, so that the suppression of the beam trigger is realized 
without the RM network.

The operating principle of the shift-jump function is shown in Fig.~\ref{fig:shift_jump}.
In the original transmitter module, 
the switching of the type sequence is allowed only when a single sequence
is finished reaching the end code or the jump code to avoid corruption of the behavior.
In the new module, the switching in the type sequence triggered by the interlock signals
is newly defined. Each type sequence has a jump table, which defines the destinations
related to the type of the interlocks. 
In the figure, we assume that Type sequence 1 is the sequence for the normal
operation, where both of the MLF and the experiment of the MR are running.
The interlock numbers of 1, 2, 3 correspond to the All-stop, MLF-inhibit, and
MR-inhibit interlocks, and 
the destinations of the interlock numbers in the jump table of Type sequence 1
are defined as to 22, 33, and 44 respectively.

If an interlock is asserted at $N$-th period, the type sequence is switched
at $(N+1)$-th period to the destination, according to the jump table. Type sequence 22 is composed
by the type codes without the beams. Type sequence 33, which correspond to MLF-inhibit,
consists of the type codes without the MLF beams.
The case is similar for the case of Type sequence 44.

Thus, the suppression of the beam trigger is realized 
by the new shift-jump function without the RM network. The jump table must be
carefully defined. The required number of type memory banks is also increased.
This is another reason why the extension of the memory bank to 1024 is implemented.

Eight interlock inputs are prepared in the new transmitter module. 
Three interlocks are already defined as described
above. Other five inputs are reserved.


\subsection{History and outlook}

The development of the next-generation timing system was started
in Japanese fiscal year (JFY) 2016, with prototyping on the Xlinx
Spartan-6 evaluation kits, SP605. The new transmitter and receiver
modules and the optical fanout were built based on insights gained
from prototyping in JFY 2017--2018.

After extensive tests,
the original transmitter module was replaced with new module
in January 2020. Also, a few replacements of the receiver stations
in the linac were made, where the noise environment is relatively
severe and stabilization of the timing system is desired.
The replacements were successful. The original receiver modules
in the facilities that receive the signal from the new transmitter
are running without issues as before and 
the new receiver modules are generating the triggers as expected.

As of September 2020, the replacements of 12 receiver stations with 28
new receiver modules in the linac and a reveiver station with 8
modules have been done. Additional 13 stations are to be replaced
in JFY 2020. Three new power supply buildings in the MR
have been constructed for the upgrades of the magnet power supplies
toward the shorter repetition period \cite{Koseki:IPAC2018-TUPAK005}. 
The new PLC receiver modules are to be fully utilized in the
new power supply buildings. The original VME receiver modules in the MR buildings
will be updated with the PLC modules depending on the budget.

The shift-jump function of the new transmitter module is under testing.
The function will be enabled for the normal user operation from
this November.

\section{Conclusion}

The J-PARC timing system is designed to fit the operation
of the accelerator complex, which has multiple beam destinations.
The original scheduled timing system has been working
without major problems, while the original system cannot
be maintained for long term, due to the discontinued
optical component. The next-generation timing system
has been therefore developed.

The new system is developed with an emphasis on
compatibility with the original system in terms of the operating principle.
The signal transfer of the system is based on the high speed serial
communication. Thanks to the CDR technology,
three timing signals, the reference trigger, clock, and type code,
are sent via a single optical cable. Particular consideration
is given to realization of the fixed latency and the CDA and 10B/8B
decoder are implemented in the FPGA fabric according to the reference \cite{giordano:tns11}.
Sufficiently low jitters of the clock and trigger are achieved. 

New features are implemented in the new transmitter module.
The shift-jump function is implemented so that the beam
trigger can be stopped without the RM network
when the interlock is asserted.

As designed to be compatible with the original system,
the installation of the new receiver modules is progressing
smoothly. The PLC receiver modules are to be installed in the new
power supply buildings of the MR. The original VME receiver modules in the MR buildings
will be replaced with the PLC modules depending on the budget.

The next-generation system is based on common serial communication technologies.
The serial communication framework can still be implemented with the newer FPGAs,
even if the current FPGA of the system is discontinued.
Therefore, the system is expected to be sustainable.

\section*{Acknowledgments}
We would like to thank the staffs of Hitachi Zosen Corporation, who
implemented the timing modules. We are grateful to the J-PARC writing
support group, which continuously encouraged us to write up this
article. Also, we would like to thank all the members of the J-PARC.  We
acknowledge the late Prof. Junsei Chiba for his contribution to 
the design of the original J-PARC timing system.

\bibliographystyle{IEEEtran}
\bibliography{tamura-bib_eng_only}

\end{document}